\documentclass[aps,prb,amsmath,amssymb,twocolumn,superscriptaddress,preprintnumbers,showpacs,longbibliography]{revtex4-2}
\pdfoutput=1
\bibpunct{[}{]}{,}{n}{}{}

\usepackage[utf8]{inputenc}
\usepackage{bm,latexsym,mathrsfs,enumerate}
\usepackage[dvipsnames]{xcolor}
\usepackage{mathtools}
\usepackage{makecell}
\usepackage{multirow}
\usepackage[breaklinks=true,unicode=true,urlcolor = blue,colorlinks = true,citecolor = blue,linkcolor = blue]{hyperref}
\usepackage{graphicx}
\usepackage{ragged2e}
\usepackage[export]{adjustbox}

\renewcommand{\vec}[1]{\bm{#1}}
\usepackage[final]{pdfpages}
\usepackage{tikz}
\makeatletter
\AtBeginDocument{\let\LS@rot\@undefined}
\makeatother
\renewcommand{\figurename}{\textbf{Figure}}

\begin{document}

\title{{\sffamily Direct evidence of a large spin-orbit coupling in the FeSe superconducting monolayer on STO}
}

\title{{\sffamily Direct probing of a large spin-orbit coupling in the FeSe superconducting monolayer on STO: Evidence for nontrivial topological states}
}

\author{\sffamily Khalil Zakeri}
\email{khalil.zakeri@kit.edu}

\affiliation{\sffamily Heisenberg Spin-dynamics Group, Physikalisches Institut, Karlsruhe Institute of Technology, Wolfgang-Gaede-Str. 1, D-76131 Karlsruhe, Germany}

\author{\sffamily Dominik Rau} \affiliation {\sffamily Heisenberg Spin-dynamics Group, Physikalisches Institut, Karlsruhe Institute of Technology, Wolfgang-Gaede-Str. 1, D-76131 Karlsruhe, Germany}

\author{\sffamily Jasmin Jandke} \affiliation {\sffamily Physikalisches Institut, Karlsruhe Institute of Technology, Wolfgang-Gaede-Str. 1, D-76131 Karlsruhe, Germany}

\author{\sffamily Fang Yang} \affiliation {\sffamily Physikalisches Institut, Karlsruhe Institute of Technology, Wolfgang-Gaede-Str. 1, D-76131 Karlsruhe, Germany}

\author{\sffamily Wulf Wulfhekel} \affiliation {\sffamily Physikalisches Institut, Karlsruhe Institute of Technology, Wolfgang-Gaede-Str. 1, D-76131 Karlsruhe, Germany}

\affiliation {\sffamily Institute for Quantum Materials and Technologies, Karlsruhe Institute of Technology, D-76344, Eggenstein-Leopoldshafen, Germany}

\author{\sffamily Christophe Berthod} \affiliation {\sffamily Department of Quantum Matter Physics, University of Geneva, 1211 Geneva, Switzerland}

\begin{abstract}

\end{abstract}
\maketitle

\onecolumngrid

 {\sffamily\textbf{\large{Abstract}}}\\
-----------------------------------------------------------------------------------------------------------------------------------------------------\\

 {\sffamily \textbf{In condensed-matter physics spin-orbit coupling (SOC) is a fundamental physical interaction, which describes how the electrons' spin couples to their orbital motion. It is the source of a vast variety of fascinating phenomena in solids such as topological phases of matter, quantum spin Hall states, and many other exotic quantum states. Although in most theoretical descriptions of the phenomenon of high-temperature superconductivity SOC has been neglected, including this interaction can, in principle, revise the  microscopic picture of superconductivity in these compounds. Not only the interaction leading to Cooper pairing but also the symmetry of the order parameter and the topological character of the involved states can be determined by SOC. Here by preforming energy-, momentum-, and spin-resolved spectroscopy experiments with an unprecedented resolution we demonstrate that while probing the dynamic charge response of the FeSe monolayer on strontium titanate, a prototype two dimensional high-temperature superconductor using slow electrons,  the scattering cross-section shows a considerable spin asymmetry. We unravel the origin of the observed spin asymmetry by developing a model in which SOC is taken into consideration. Our analysis indicates that SOC in this two dimensional superconductor is rather strong.
 We anticipate that such a strong SOC can have several serious consequences on the electronic structures and can lead to the formation of topological states. Moreover, a sizable SOC can compete with other pairing scenarios and is crucial for the mechanism of high-temperature superconductivity.}}\\
--------------------------------------------------------------------------------------------------------------------------------------------------------


\section{Introduction}\label{Sec:Intro}

The fundamental interaction describing the microscopic coupling mechanism between the spin and orbital degrees of freedom of electrons in solids is the so-called spin-orbit coupling (SOC) \cite{Galitski2013,Winkler2003}. This interaction, which is a relativistic effect, is an essential ingredient for describing many emergent phenomena observed in condensed-matter systems. For instance, a large SOC in combination with other symmetry aspects can lead to the appearance of topological phases in solids.
Examples of this kind are the topological insulators, where a large SOC leads to the formation of the topologically protected surface states and spin momentum locking \cite{Moore2010,Hasan2010,Qi2011,Shen2017,He2018}.
Likewise in magnetically ordered solids SOC in the absence of inversion symmetry can result in the formation of topologically protected spin textures in the form of chiral domain walls, skyrmions, antiskyrmions, hopfions, etc. \cite{Zang2018}.

In order to figure out whether or not a material exhibits topological electronic states and to which topological classes these states belong, one requires to quantify the strength of SOC. Assuming that the symmetry considerations are fulfilled, the presence of a sufficiently large SOC would, in principle, give rise to the formation of nontrivial topological states in the system. Although the phenomenon of high temperature superconductivity is, by itself, a fascinating phenomenon, combined with topological aspects of matter it would lead to an even more exotic state of matter  e.g., topological superconductivity and the formation of the Majorana states \cite{Leijnse2012,Beenakker2013,Sato2017}. These states which obey non-Abelian statistics can be used to realize topological quantum computers \cite{Nayak2008}. In most of the proposals for realizing these interesting concepts it is suggested to attach a low-dimensional superconductor to a topological material or semiconductor heterostructures with a large SOC \cite{Beenakker2013}. However, under some circumstances if SOC in a low-dimensional superconductor is sufficiently large, one expects to observe topological states in a single material \cite{Hao2018}. An ideal candidate for such an observation would be a single layer of FeSe grown on SrTiO$_3$(001), an ideal two-dimensional high temperature superconductor (HTSC) \cite{Wang2012a,Liu2012,Tan2013,He2013,Bozovic2014,Ge2014,Lee2014}.
In the case of ultrathin films the inversion symmetry  in the direction perpendicular to the surface is broken. A large SOC together with the broken inversion symmetry can provide the necessary fundamental basis required for the observation of topological states in the system \cite{Kang2016, Hao2015}. Hence a direct probing of SOC in this class of materials is essential in connection with the possibility of the formation of topological states. Unfortunately, the strength of SOC in such two-dimensional superconductors is hitherto fully unknown.

Irrespective of  the importance of SOC for the topological superconductivity, the presence of this interaction is of prime importance to understand the underlying physics of HTSC in general \cite{Borisenko2015,Kang2016}. In most theoretical approaches describing the microscopic mechanism of superconductivity and Cooper pairing of electrons SOC is assumed to be very small and, therefore, has been neglected. There are only a few theoretical models which include SOC in bulk HTSCs, showing that the presence of this interaction is essential in the determination of the symmetry of the order parameter as well as the electronic states involved in superconductivity \cite{Kang2016,Borisenko2015}. Generally the impact of SOC becomes increasingly important when reducing the systems' dimensionality. This is due to the emergence of new symmetry aspects in low-dimensional solids. Surprisingly, so far no direct signature of SOC and its impact in ultrathin (two-dimensional) HTSCs have been reported experimentally. A direct measure of SOC would, therefore, be extremely valuable in the context of microscopic physical mechanism behind high temperature superconductivity in low-dimensional HTSCs.

Here by performing high resolution spectroscopy of spin-polarized slow electrons on  epitaxial FeSe monolayers grown on Nb-doped strontium titanate SrTiO$_3$(001), a prototypical two-dimensional superconductor, we demonstrate that the frequency and momentum dependent scattering cross-section depends strongly on the spin of the incoming electron. A careful analysis of the spectra reveals that the observed effect is due to the presence of a considerably large SOC in this system. Such a large SOC together with other symmetry aspects provides the required ingredients for the formation of topologically nontrivial states and would shed light on the mysterious origin of superconductivity in this system.

\section{Results}\label{Sec:results}

The epitaxial FeSe monolayer was grown by molecular beam epitaxy on Nb-doped SrTiO$_3$(001) (hereafter STO). The dynamic charge response of the system was probed by means of spin-polarized high-resolution electron energy-loss spectroscopy (SPHREELS) (see Sec.~\ref{Sec:Exp} of Materials and Methods for details on the substrate preparation, film growth and SPHREELS experiments). The scattering geometry is sketched in Fig.~\ref{Fig:Spectra}\textbf{a}. The scattering plane was chosen to be parallel to the [100]-direction of STO(001), as indicated in Fig.~\ref{Fig:Spectra}\textbf{b} and \textbf{c}. This would allow probing the dynamic response of the system along the high symmetry $\bar{\Gamma}$--$\bar{\rm X}$ direction of the surface Brillouin zone (SBZ). In order to be sensitive to the spin-dependent effects associated with the broken inversion symmetry in the direction perpendicular to the surface, we used a longitudinally spin-polarized electron beam with the spin orientation being parallel and antiparallel to the scattering's plane normal vector $\vec{n}$ (see Supplementary Note~1 for an explanation). These incoming spin states are denoted by $|+\rangle$ and $|-\rangle$, respectively. 
Figure~\ref{Fig:Spectra}\textbf{d} shows the spin-resolved spectra recorded in the superconducting state of the sample and using an incident electron beam energy $E_i=4.07$~eV. The spectra were recorded at the $\bar{\Gamma}$--point of SBZ. Beside the so-called zero loss peak at the energy-loss  $\hbar\omega=0$, one observes several features as a result of the excitation of several collective modes. The peaks with lower intensity at $\hbar\omega=11.8$, 20.5, 24.8 and 36.7~meV represent the various phonon modes of the FeSe film itself \cite{Zakeri2017,Zakeri2018,Zhang2018}. More obviously the so-called Fuchs-Kliewer (FK) phonon modes of the underlying STO substrate can also be recognized at the loss energies $\hbar\omega=59.3$ and 94.5~meV \cite{Zhang2016,Jandke2019}.

\setcounter{figure}{0}
\makeatletter
\renewcommand{\figurename}{\textbf{Figure}}
\renewcommand{\thefigure}{\textbf{\@arabic\c@figure}}
\makeatother
\begin{figure*}[t!]
	\centering
	\includegraphics[width=0.7\columnwidth]{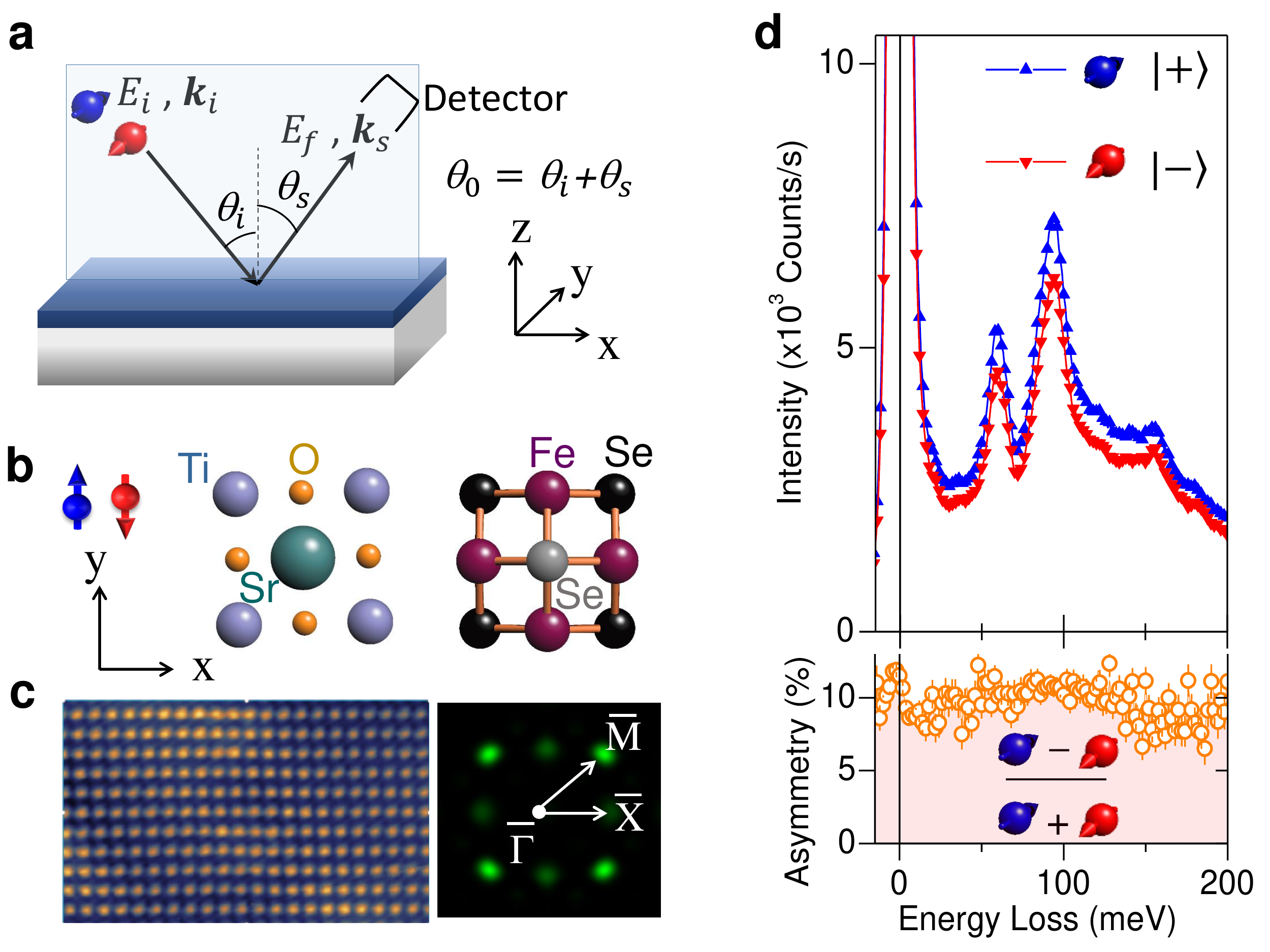}
	\caption{\textbf{Evidence of a large SOC in FeSe ML on STO.}  \textbf{a} The scattering geometry used for probing the dynamic charge response. The electrons are represented by the blue and red balls. Their spin in the laboratory frame is shown by the red and blue arrows. The incident energy and wavevector are denoted by $E_i$ and $\vec{k}_i$, respectively. The energy and the wavevector after the scattering event are given by $E_f$ and $\vec{k}_s$, respectively. The laboratory frame is depicted by black arrows with x, y, and z labels. The incident and outgoing angles are called $\theta_i$ and $\theta_s$. The total scattering angle is  $\theta_0$ and was set to 80$^\circ$. \textbf{b} The top view of  STO(001) and the FeSe(001) film. The spin polarization of the beam is either parallel or antiparallel to the y-axis that is the [010]-direction of the STO(001) surface. These spin states are called  $|+\rangle$ and $|-\rangle$, respectively. \textbf{c} Atomically resolved scanning tunneling microscopy topography image of the FeSe surface, showing the atomic resolution of the topmost Se atoms, indicated by the gray balls in \textbf{b}. The field of view is $7\times4.5$~nm$^2$. The constant current topography image was recorded at $T=0.9$~K and using a tunneling current of 180~pA and a bias voltage of 1.0~V. The corresponding reciprocal lattice is shown in the right side. The $\bar{\Gamma}$--point represents the SBZ center and the $\bar{\rm X}$ and $\bar{\rm M}$ points represent the edges of SBZ. \textbf{d} Blue upward and red downward triangles represent the experimental spectra recorded for the spin of the incoming beam being parallel and antiparllel to the y-axis, respectively. The open circles denote the spin asymmetry $A=\left(I_{|+\rangle}-I_{|-\rangle}\right)/\left(I_{|+\rangle}+I_{|-\rangle}\right)$. The spectra were recorded at the specular geometry i.e., the wavevector of $q=0$, at the $\bar{\Gamma}$--point. The error bars represent the statistical uncertainties.}
	\label{Fig:Spectra}
\end{figure*}

Generally the spectral function $\mathcal{S}(q, \omega)$ measured by SPHREELS directly reflects the dynamic response of the collective charge excitations in the system. This quantity is proportional to the imaginary part of the dynamic charge susceptibility $\mathfrak{Im}\chi(q, \omega)$ \cite{Vig2017,Husain2019,Zakeri2021}. The electrons are scattered by the total charge distribution of the sample and, hence, the scattering intensity carries information regarding collective excitations of the lattice i.e, phonons, collective electronic excitations i.e., plasmons and any type of excitation representing a hybrid mode of these two \cite{Zakeri2021}.
The most interesting observation here is that  $\mathcal{S}(q, \omega)$ depends strongly on the spin. The spin asymmetry defined as $A=\left(I_{|+\rangle}-I_{|-\rangle}\right)/\left(I_{|+\rangle}+I_{|-\rangle}\right)$ is shown in the lower part of Fig.~\ref{Fig:Spectra}\textbf{d}. Here $I_{|+\rangle}$ and $I_{|-\rangle}$ denote the intensity of the scattered electrons when the incoming electron's spin is parallel and antiparallel to $\vec{n}$, respectively.

\begin{figure*}[t!]
	\centering
	\includegraphics[width=0.99\columnwidth]{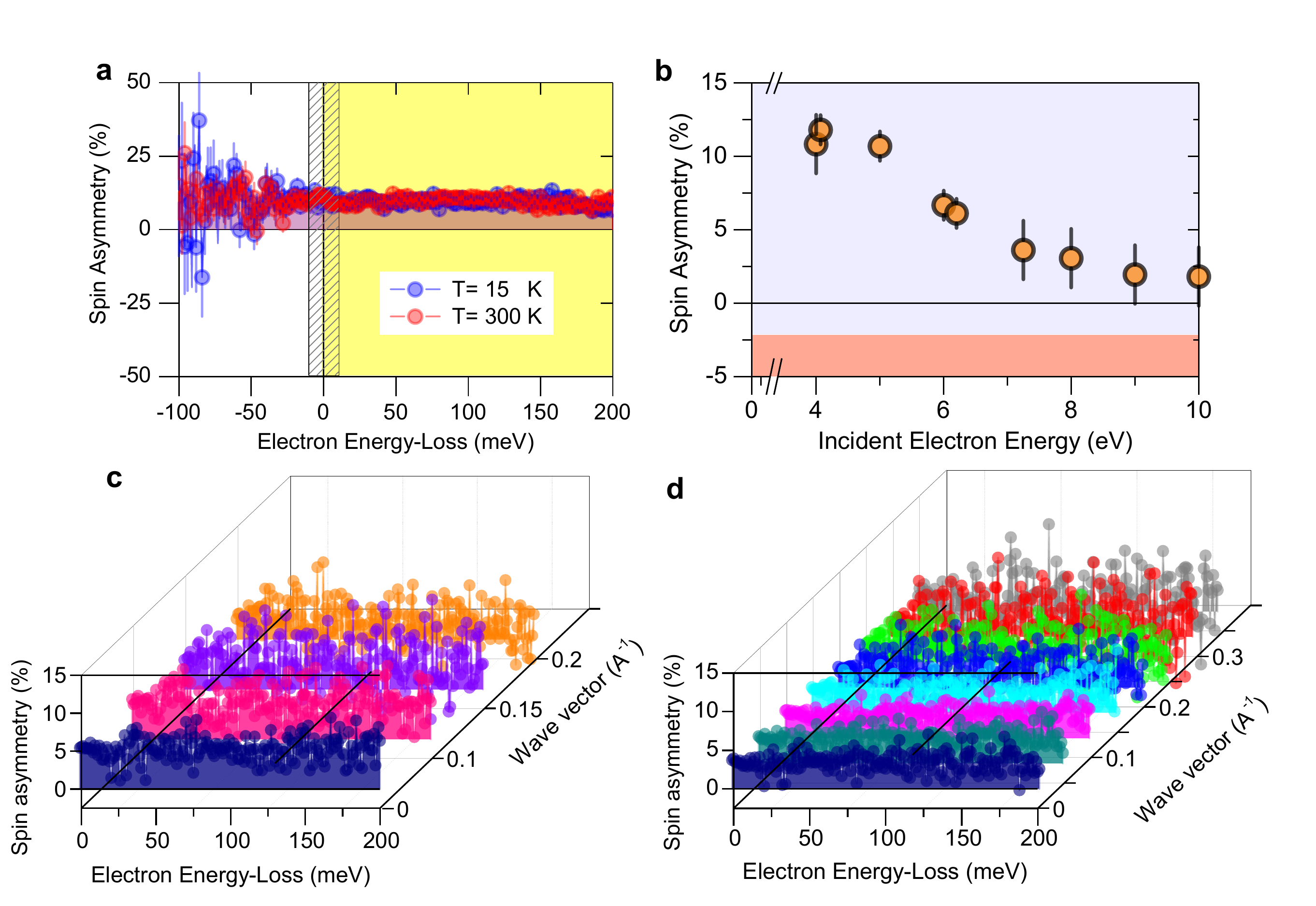}
	\caption{\textbf{SOC as the origin of the observed spin asymmetry.} \textbf{a} Spin asymmetry measured below ($T=15$~K) and above ($T=300$~K) the superconducting transition temperature. The data are recorded at an incident beam energy of 4.07~eV and at the specular geometry ($q=0$). The error bars represent the statistical uncertainties. \textbf{b} Spin asymmetry as a function of the incident beam energy. The error bars represent the systematic uncertainties.  \textbf{c},\textbf{d} The pattern of the spin asymmetry over the energy loss for different values of wavevector along the $\bar{\Gamma}$--$\bar{\rm{X}}$ direction. The graphs represent the spin asymmetry of the spectral function $\mathcal{S}(q, \omega)$. The data shown in \textbf{c} are recorded at an incident beam energy of 6.0~eV and those shown in \textbf{d} are recorded at 7.25~eV. }
	\label{Fig:TandEdep}
\end{figure*}

In order to shed light on the origin of the observed spin asymmetry, its dependency on the physical variables e.g., temperature, incident energy, and wavevector transfer $q$ was measured and the results are summarized in Fig.~\ref{Fig:TandEdep}. Data presented in Fig.~\ref{Fig:TandEdep}\textbf{a} clearly demonstrate that the spin asymmetry does not depend on temperature. In both superconducting and normal states one observes a value as large as 11\%. This fact indicates that the spin asymmetry is not related to the superconducting (or magnetic) phase transition and is due to the intrinsic SOC of the system.  Next we check the dependence of the spin asymmetry on the incident beam energy $E_i$. Generally for very low incident energies ($E_i<3$~eV) the intensity may be influenced by the space charge effects. On the other hand for incident energies higher than 12~eV the intensity is determined by the multiple scattering and electron diffraction processes (the so-called low-energy electron diffraction or LEED states). Hence, the relevant energy window would be between 3 and 11~eV. Such data are presented in Fig.~\ref{Fig:TandEdep}\textbf{b}. For this set of measurements first the incident and scattered beam angles were fixed to $\theta_i=\theta_s=40^{\circ}$ (see Fig.~\ref{Fig:Spectra}\textbf{a}). The incident beam energy $E_i$ was precisely defined and the electrons with the final energy $E_f=E_i \pm \delta E$ were collected. Here $\delta E$ represents the energy width of the elastic scattering (the hatched area in Fig.~\ref{Fig:TandEdep}\textbf{a}). In order to make sure that all the elastically scattered electrons are collected, we recorded the intensity for $\delta E=8$~meV (this value is two times the energy resolution). The spin asymmetry shows a strong dependence on the incident beam energy and exhibits a maximum near 4~eV. As the next physical variable we check the dependence of the spin asymmetry on $q$. Spectra recorded for different values of $q$ near the zone center (in the vicinity of the $\bar{\Gamma}$--point) indicate that the spin asymmetry does not depend on $q$. This is demonstrated in Figs.~\ref{Fig:TandEdep}\textbf{c} and \textbf{d}, where the spin asymmetry recorded for different values of $q$ is presented. The data shown in Fig.~\ref{Fig:TandEdep}\textbf{c} were recorded with an incident beam energy of $E_i=6.0$~eV and those in Fig.~\ref{Fig:TandEdep}\textbf{d} were recorded with $E_i=7.25$~eV. A careful inspection of the data shown in Figs.~\ref{Fig:TandEdep}\textbf{c} and \textbf{d} indicates that although the spin asymmetry depends strongly on $E_i$,  it does not depend on $q$.  The strong $E_i$-dependence of spin asymmetry and its $q$-independence is an unambiguous evidence that the observed spin asymmetry is originating from a substantially large SOC at the surface (see the discussion below).

\section{Discussion}\label{Sec:discussions}

It is well-known that when a beam of spin-polarized slow electrons is scattered from a free atom with a large atomic number, and consequently a large SOC, the scattering cross-section can be spin dependent \cite{Mott1929a,Kessler1985}. The effect is understood based on the fact that due to the relativistic effects the electrons with different spins feel different scattering potentials, while scattered off the atom. The apparent different scattering potentials are the direct consequence of the relativistic motion of electrons in the vicinity of the atom \cite{Kessler1985}. The same phenomenon has also been observed when such a  beam is scattered from a surface i.e., an array of atoms ordered in a two-dimensional fashion  \cite{Kessler1985,Kirschner1985,Feder1986, Wang1979}. The effect is attributed to the intrinsic SOC of the involved atomic orbitals. As a consequence of the broken translation symmetry at the surface, the largest effect is observed when the spin of the incoming electron beam is parallel and antiparallel to the scattering's plane normal vector $\vec{n}$ and hence the spin asymmetry is maximum in this case.

In a simplified version the spin dependent elastic scattering cross-section $dS/d\Omega$ may be written as
\begin{equation}
	\dfrac{d S}{d \Omega}=\dfrac{d S_0}{d \Omega} \left\lbrace 1+ \sum_\gamma \delta_{l_\gamma}   \left[\vec{\xi}\cdot\left( \vec{\hat{{k}}}_i \times \vec{\hat{{k}}}_s\right)\right] \right\rbrace,
	\label{Eq:elasticcrosssection}
\end{equation}
where $dS_0/d\Omega$ is the scattering cross-section without considering SOC, $ \sum_\gamma \delta_{l_\gamma}$ represents the sum over all the possible scattering phase shifts $\delta_{l_\gamma}$ between partial waves with total angular momenta between $l-1/2$ and $l+1/2$ ($l$ is the orbital quantum number of the involved atomic orbitals), $\vec{\xi}$ represents the spin of the incoming beam, $\vec{\hat{{k}}}_i$ and $\vec{\hat{{k}}}_s$ denote the unit vectors of the momentum of the incoming and scattered beam, respectively \cite{Kessler1985,Feder1986,Sushkov2013}. Note that Eq.~(\ref{Eq:elasticcrosssection}) is valid for the energies above the so-called centrifugal barrier, which is on the order of 1--3~eV.

Equation~(\ref{Eq:elasticcrosssection}) indicates that the value of the spin asymmetry depends only on $\sum_\gamma \delta_{l_\gamma}$, which, in turn, depends on the incident energy $E_i$. The spin asymmetry does not depend on $q$, since different values of $q$ are achieved by changing the scattering geometry, keeping the total scattering angle $\theta_0=\theta_i+\theta_s=80^{\circ}$ fixed. For a longitudinally spin-polarized beam, since the asymmetry is given by the vector product $\vec{\hat{{k}}}_i \times \vec{\hat{{k}}}_s$, it does not change as long as the angle between $\vec{\hat{{k}}}_i$ and $\vec{\hat{{k}}}_s$ is unchanged. More importantly the maximum asymmetry is expected when the second term in Eq.~(\ref{Eq:elasticcrosssection}) is maximum. Such a condition is realized for a longitudinally spin-polarized beam in which the direction of spin polarization vector is parallel or antiparallel to the cross product $\vec{\hat{{k}}}_i \times \vec{\hat{{k}}}_s$ or the scattering's plane normal vector  $\vec{n}$ (see Supplementary Note~1 for  details).

\begin{figure*}[t!]
	\centering
	\includegraphics[width=0.45\columnwidth]{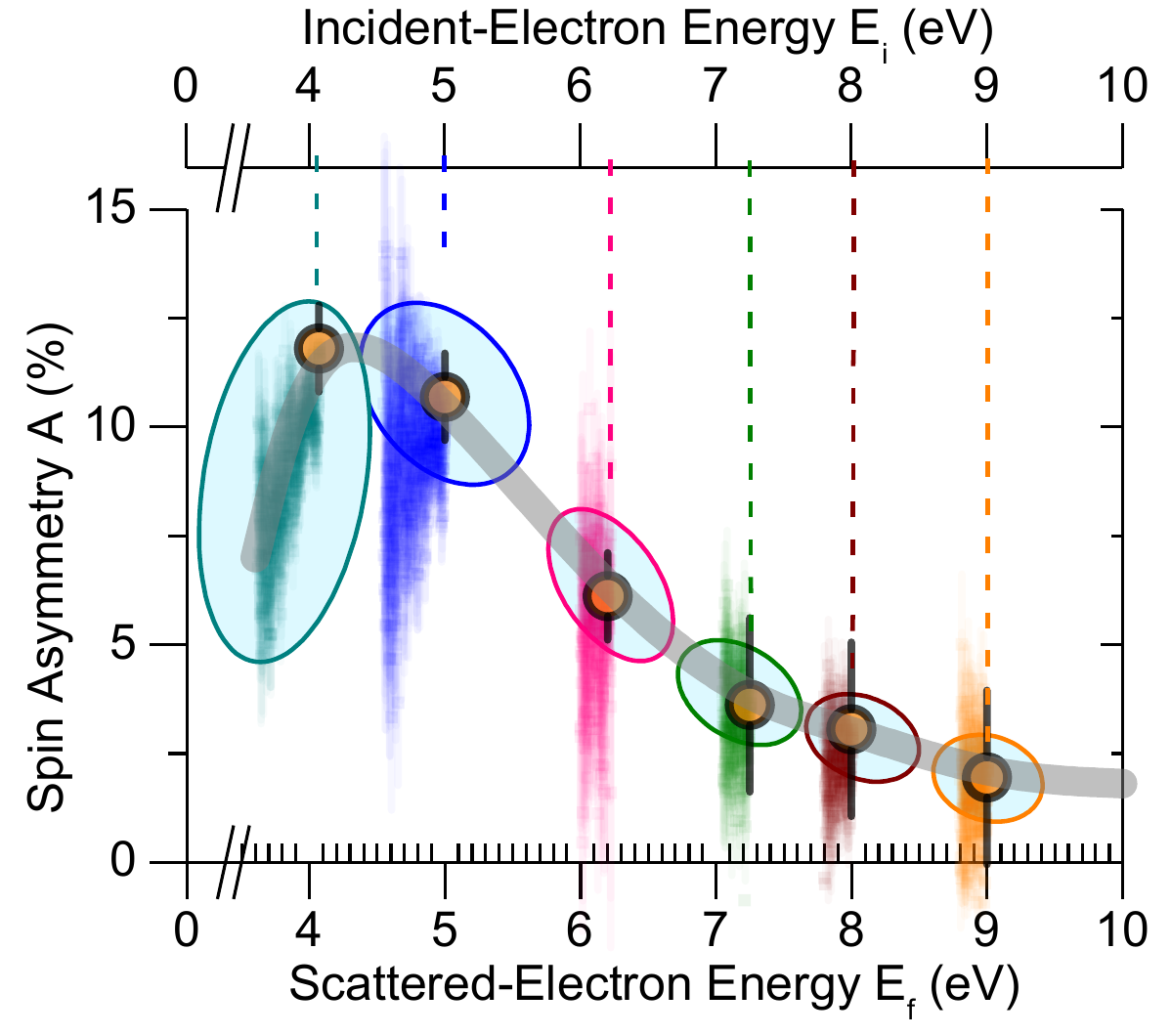}
	\caption{\textbf{Dependence of the  spin asymmetry on the energy of the scattered beam}. Spin asymmetry $A$ versus scattered beam energy $E_f$, when the incident beam energy $E_i$ is kept constant. For each set of data the energy of the incident beam $E_i$ is also shown using the top axis; $E_i=4$ (teal color), 5 (blue color), 6.2 (red color), 7.3 (green color), 8 (brown color) and 9 (orange color) eV.  The filled circles represent the elastic scattering i.e., cases in which $E_i=E_f$. The gray thick curve is a guide to the eye and describes the $E_i$-dependence of $A$ shown in Fig.~\ref{Fig:TandEdep}\textbf{b}. The data indicate that the $E_f$-dependence of $A$ is the same as its $E_i$-dependence. The error bars represent the systematic uncertainties.}
	\label{Fig:Lossdep}
\end{figure*}

 When electrons are scattered from a surface the scattering process near the specular angle is governed by the dipolar scattering mechanism \cite{Ibach1982}. In the so-called dipolar lobe, the electrons interact with the charge density fluctuations of the sample. The interaction is of the long range dipolar (Coulomb) nature. Hence, not only the charge density fluctuations near the surface region but also those located far below the surface can be observed in the spectra, depending on the kinetic energy of the incoming beam \cite{Ritz1984,Schaich1984,Lueth1988}. Since we are interested in the properties of the FeSe/STO interface, we use electrons with kinetic energies as low as 4 up to 10~eV. This choice of incident energy is also important to, on the one hand keep the incident electron energy above the centrifugal barrier and, on the other hand, avoid multiple scattering processes, whose presence would add to the complexity of the problem when one aims to understand the spin dependent scattering cross-section.

Beside the elastic part of the spectrum, an even more interesting part is the energy-loss region, where the collective excitations of the system show up. We, therefore, carefully analyzed the spin asymmetry as a function of electrons' energy in the final state after scattering $E_f$, while keeping the energy of the incident electrons constant. The data are presented in Fig.~\ref{Fig:Lossdep}. For this set of measurements we first optimized the incoming beam at a given energy $E_i$ and probed the spin asymmetry of the elastic scattering at $E_f=E_i\pm \delta E$. The results are shown by the filled circles in Fig.~\ref{Fig:Lossdep}. Then for each value of $E_i$ we probed the spin asymmetry as a function of $E_f$ over an energy range for which we could obtain reasonable count rates.  The experiment was performed for various values between 4 and 9~eV. The results clearly demonstrate that the spin asymmetry versus scattered beam energy follows the same trend as that of the elastic scattering for different energies. For example for the electrons with the incident energy of $E_i=4$~eV the asymmetry decreases when decreasing $E_f$ from 4 to 3.5~eV. However, for the higher incident energies the spin asymmetry gradually increases when moving towards lower values of $E_f$. This means that the observed spin asymmetry in the energy loss region has the same origin as that of the elastic scattering. In simple words electrons contributing to the surface loss processes are then affected by the SOC potential before they are finally scattered out.  In the limit of small energy losses ($\hbar\omega=E_i-E_f \ll E_i$) the spin asymmetry is almost entirely determined by the incident energy $E_i$.

\begin{figure*}[t!]
\centering
\includegraphics[width=0.99\columnwidth]{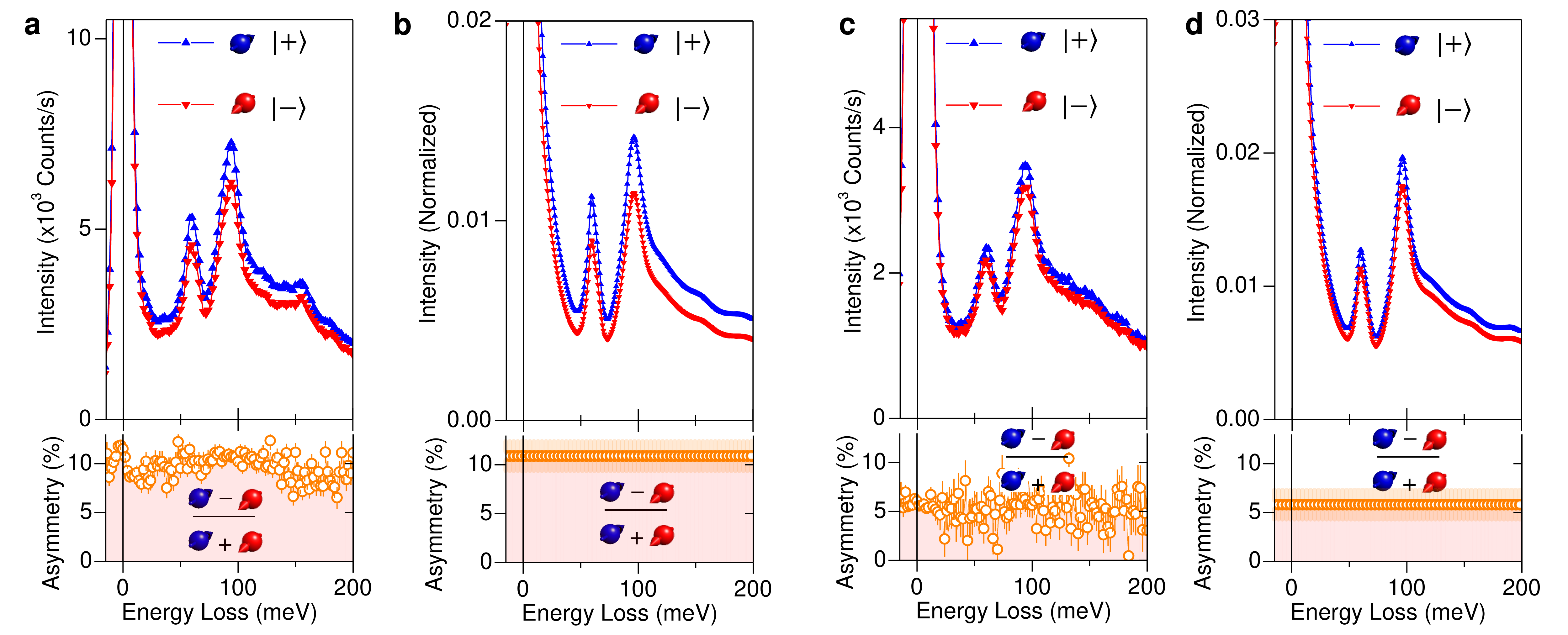}
\caption{\textbf{Spin dependence of dynamic charge response.} \textbf{a} and \textbf{b} The measured and simulated SPHREEL spectra at an incident electron energy of $E_i=4.07$~eV.  \textbf{c} and \textbf{d} The measured and simulated SPHREEL spectra at an incident electron energy of $E_i=6.0$~eV. Blue upward and red downward triangles represent the spectra recorded or simulated for the incoming spin states of $|+\rangle$ and $|-\rangle$ (spin polarization parallel and antiparallel to the y-axis in  Fig.~\ref{Fig:Spectra}\textbf{a}), respectively. The open circles denote the spin asymmetry  $A=\left(I_{|+\rangle}-I_{|-\rangle}\right)/\left(I_{|+\rangle}+I_{|-\rangle}\right)$. All the experimental spectra are recorded at $T=15$~K in the superconducting state of the sample and at the specular geometry i.e., the wavevector of $q=0$, at the $\bar{\Gamma}$--point. The error bars in spin asymmetry represent the statistical uncertainties.}
	\label{Fig:SpectraSimulation}
\end{figure*}

Yet the question whether or not the dynamic charge response depends on the spin of the incident electron remains unanswered. In order to answer this question and shed light onto the origin of the observed spin asymmetry we developed a model to simulate the SPHREEL spectra. The simulation is based on the scattering theory of spin-polarized slow electrons from a surface with a nonnegligible SOC. The theory is an extension of the original theory of Evan and Mills \cite{Evans1972,Mills1975,Ibach1982}, which describes the scattering cross-section of an unpolarized electron beam. In our modeling of the scattering event and in the calculation of the scattering cross-section we further account for SOC and spin dependent electron reflection (in addition to the Hartree or Coulomb scattering, see Sec.~\ref{Sec:Theo} of Materials and Methods for details). Our theory indicates that for small values of momentum transfers the observed spin dependent asymmetry can be explained based on the spin dependence of electron reflectivity. Hence, the dynamic charge response itself is rather spin independent. The observed spin asymmetry is almost entirely due to the spin dependent electron reflection from the surface, which in turn is a result of SOC. In order to verify this hypothesis we simulated the SPHREEL spectra using the values of spin dependent reflection coefficient measured for the elastic scattering (data shown in Fig.~\ref{Fig:TandEdep}\textbf{b}). The results of simulation for two different values of $E_i$ are shown in Fig.~\ref{Fig:SpectraSimulation} together with the experimental spectra. Our theory is able to perfectly reproduce the experimental spectra.  In the simulation we considered a system composed of an atomic layer of FeSe (Se-Fe-Se trilayer structure) on $17$ unit cells of charge free insulating STO(001) on top of a semi-infinite Nb-STO(001). In this model the Fe plane in FeSe ML is placed at $d_{\rm{FeSe}}=0.43$~nm above the insulating STO(001) surface \cite{Peng2020}. Only in this way both the peak position and amplitude of the excitations associated with the FK modes agree with those measured experimentally.  The discovery of a charge depletion layer below the FeSe layer has been discussed in details in Ref.~\cite{Zakeri2022}. Both simulations and experiments reveal that the higher harmonics of the principal FK modes are strongly suppressed. This is mainly due to the presence of the free carriers in FeSe ML as well as in the interior part of the substrate, below the depletion region. As discussed in Ref.~\cite{Zakeri2022} the presence of a charge depletion layer, which is due to a considerably large charge transfer from the Nb-doped STO to the FeSe film, has several serious consequences on the system. One of the consequences is that it generates a rather large electric field and band bending near and below the interface. As a rule of thumb one may simply divide the value of the total band  bending, probed by the experiment (2.1~V), by the depletion layer thickness (6.5~nm). This results in an electric field on the order of  0.3~GV/m. Such a large electric field greatly influences the electronic states in FeSe ML and boosts SOC at and near the surface region. The effect is very similar to the Rashba and Dresselhaus effects observed for semiconductor quantum-well states and two-dimensional electron gasses formed at the semiconductor  surfaces and interfaces \cite{Winkler2003}.

Experiments performed on bulk FeSe using neutron scattering \cite{Ma2017} and on other Fe-based superconductors using angle-resolved photoemission spectroscopy (ARPES) \cite{Borisenko2015} have shown that the Fe atoms possess a nonnegligible SOC. Since in the present case the FeSe monolayer is subject to a rather strong electric field as a result of the charge transfer and the dielectric depletion layer, we conclude that the observed large SOC in this system is, therefore, an additive effect. It includes the intrinsic SOC of FeSe  ML as well as the electric field induced SOC, as discussed above. In order to provide an estimation of the strength of SOC one may compare the results to that of the W(110) surface. In the case of W(110) the largest spin asymmetry was observed for $E_i=4.7$~eV and was about 64\% with an electron beam with a polarization of about 72\% (this would mean a corrected value of about $A=62\%/72\% \approx 86\%$). For the case of Cu(001) the spin asymmetry is negligible (it is below 0.5\%) over the same range of incident energy. Comparing these results to the value of about 8\% for the same beam (or 11\% after correction), one may conclude that SOC in FeSe ML on STO is by a factor of about 8 smaller than that in W(110). The value is, however, sufficiently large to cause topological states \cite{Hao2014,Hao2015,Wang2016a}.

The conditions under which FeSe ML on STO can exhibit topological characters have theoretically been examined in Refs.~\cite{Hao2014,Wang2016a}. It has been discussed that the appearance of a topological phase in this system is associated with (i) a trivial bandgap in the electronic band structure near the $\bar{\rm{M}}$-point, (ii) the parity-broken coupling at the interface, and (iii) a sufficiently large SOC \cite{Hao2014}. The presence of the bandgap near the $\bar{\rm{M}}$--point has already been well established by several ARPES experiments \cite{Lee2014,Wang2016,Tan2013,Zhang2017,Faeth2021,Rademaker2021,Liu2021}. The parity-broken coupling at the interface does exist in the system and, in addition, can be tuned by the interface engineering. Our results indicate that SOC in this system is large and hence one would, in principle, be able to realize the nontrivial topological states in the system. This would open an avenue for investigation of topological superconductivity and Majorana states in a model system with a rather high $T_c$. Evidence of topological states in FeSe ML has been reported by scanning tunneling spectroscopy \cite{Wang2016a}. However, the attribution of the observed peaks in the tunneling spectra to the topological states has entirely been based on the first-principles calculations, in which SOC was taken into account. An important piece of this puzzle i.e., a direct experimental probe of SOC has been missing. This is now provided by our experimental results.

In a similar system, namely bulk FeSe$_x$Te$_{1-x}$, the presence of topological surface superconductivity has been proposed to originate from a band inversion of the $p_z$ and $d_{xz}$ bulk bands. The resulting topological surface states have, therefore, been described to be Z$_2$ invariant (and hence be protected by time-reversal symmetry) \cite{Xu2016,Wu2016,Zhang2018a}. ARPES experiments have revealed that in addition to the superconducting gap in the vicinity of the Fermi level the surface states become gapped several mili-electron-volts below the Fermi level \cite{Peng2019,Rameau2019,Zaki2021}. This gap opening, which emerges together with the superconducting phase has been  attributed to the appearance of a weak ferromagnetism at the surface. It has been discussed that the appearance of ferromagnetism breaks the time reversal symmetry, which is in contradiction to the picture of topological superconductivity induced by the band inversion. Recently, it has been discussed that the emergence of topological superconductivity on surfaces can be the consequence of the interplay between the $s_{\pm}$-wave symmetry of the superconducting order parameter, a Rashba type SOC and the emergent surface magnetism \cite{Mascot2022}. This mechanism explains several experimentally observed fingerprints of topological superconductivity at the surface of bulk FeSeTe superconductor  \cite{Xu2016,Wang2018,Zhang2018a,Rameau2019,Zhu2020,Wu2021} and thin films \cite{Peng2019,Chen2019}. The proposed mechanism can also be important to either understand or to realize topological edge superconductivity in the FeSe ML on STO, by introducing magnetic impurities. However, such a realization would require a more detailed study of the involved states by means of ARPES  or similar techniques \cite{Wang2016}.

Regarding the origin of superconductivity in FeSe ML on STO the main attention is paid to the so-called replica bands observed in ARPES experiments. They have been considered as an indication of a phonon-mediated superconductivity in this system \cite{Lee2014,Wang2016,Tan2013,Zhang2017,Faeth2021,Rademaker2021,Liu2021}. This suggestion is merely based on probing the quasiparticle band dispersions. However, no solid evidence, indicating a strong electron-phonon coupling, has been reported by probing the phononic excitations.
On the other hand it has also been discussed that the pairing mechanism can be of unconventional nature and be mediated by spin fluctuations \cite{Graser2009,Linscheid2016}. The signature of  such a coupling has experimentally been observed by means of tunneling spectroscopy experiments \cite{Jandke2019, Liu2019}. Moreover, it has been suggested that a cooperative effect of several bosonic excitations may be the main reason for such a high transition temperature of this system \cite{Song2019,Schrodi2020a,Rademaker2021}.
Beside the mechanism discussed above, it has theoretically been shown that considering both, the effect of nematic fluctuations and SOC in the absence of inversion symmetry, a  $s-$wave state is favored \cite{Kang2016}. More importantly, if SOC and the broken inversion symmetry are strong enough so that they overcome the mismatch between the electron pockets near the $\bar{\rm{M}}$-point, the gap function measured in the experiment can be well reproduced \cite{Kang2016}. Our result, indicating a large SOC in FeSe/STO, can now provide a clue to  understand the gap function and the nature of gap opening in great details. Together with the observation of spin fluctuations, we have reported earlier \cite{Jandke2019}, we conclude that the superconductivity in the FeSe monolayer on STO is of $s_{\pm}$ character. Our results provide
critical insights into the symmetry of the paring and the origin of superconductivity in this system.

\section{Conclusions}\label{Sec:conclutions}

While probing the dynamic charge response of the FeSe superconducting ML on STO by means of SPHREELS, we observed that the scattering cross-section is strongly spin dependent. The observed spin asymmetry is attributed to a large SOC at and near the surface region. This large SOC, which very likely originates from the $d$ orbitals of the Fe atoms \cite{Kang2016} and the presence of a dielectric depletion layer below the FeSe ML, has several consequences on the properties of the system. One of the important consequences is the formation of topological states. The depletion layer is a result of charge transfer from the Nb-STO into the FeSe layer, which leads to a band bending and the generation of an electric field at the interface.

Since the charge transfer, the formation of a depletion layer at the interface and the associated electric field is a general phenomenon, we anticipate that  the observed effect is also general and shall be observed  for other Fe-based superconducting monolayers brought in contact with STO or other dielectric oxides. This would allow an interfacial engineering of the superconducting states by growing ML iron chalcogenides or other HTSC on dielectric oxides. Realization of topological states in superconducting MLs would provide a platform for investigation and realization of Majorana states in, structurally, very simple systems.

In addition to the facts mentioned above, our results suggest that the superconductivity in FeSe ML is very likely of $s_{\pm}$ nature and, therefore, shed light on the long-standing question regarding the nature of superconducting order parameter and the pairing symmetry in this system.

\section{Methods}\label{Sec:Methods}
\subsection{Experiments}\label{Sec:Exp}

\subsubsection{Sample preparation and characterization }
All the sample growth and characterizations were performed under ultrahigh vacuum conditions. FeSe ML was grown by means of molecular beam epitaxy technique on the Nb-doped STO(001) substrates in a separate chamber \cite{Jandke2019,Li2014}. The Nb-doping level was 0.6\%. Prior to the film growth, the substrate was annealed at temperatures up to about 1000~$^{\circ}$C and was then etched by a selenium flux for 20 minutes. The sample was kept at the elevated temperature for 30 minutes and then was gradually cooled down to 480~$^{\circ}$C. Fe and Se were co-deposited with a growth rate of 0.059~ML/min at 480~$^{\circ}$C at a flux ratio of Fe:Se $\equiv$ 1:10.  The sample was post annealed at 500~$^{\circ}$C right after the deposition for several hours to ensure a good morphological quality. This also ensures desorption of the residual Se atoms on the surface. The film growth was monitored by reflection high energy electron diffraction.

For the further measurements, the sample was transferred using an ultrahigh vacuum suitcase to the scanning tunneling microscopy and spin-polarized high-resolution electron energy-loss spectroscopy chambers.

The morphological and electronic properties were investigated by means of scanning tunneling microscopy \cite{Jandke2019}. Our samples show a superconducting gap of about $\Delta=11\pm3$~meV at a temperature of about 0.9~K.

\subsubsection{Spin-polarized high-resolution electron energy-loss spectroscopy}

The dynamic response of the samples was investigated by means of spin-polarized high-resolution electron energy-loss spectroscopy \cite{Zakeri2014}.

A spin-polarized monochromatic electron beam with an energy resolution between 4 and 11~meV  was used \cite{Zakeri2013,Zakeri2014}. The spin-polarized electron beam is generated by photoemission from a strained GaAsP photocathode. In order to observe maximum spin asymmetry according to Eq.~(\ref{Eq:elasticcrosssection}), a longitudinally spin-polarized beam was used, meaning that the  polarization vector of the incoming electron beam was either parallel or antiparallel to the scattering's plane normal vector $\vec{\hat{n}}$.  For the parallel case we call these incoming spin states  $|+\rangle$ and for the antiparallel case we call them $|-\rangle$. The degree of the beam polarization was estimated by performing spin-polarized elastic reflectivity from a W(110) single crystal, resulting in a value of about $72\pm5$\%. The total scattering angle i.e., the angle between the incident and the scattered beam was kept constant ($\theta_0=80^{\circ}$). The scattering intensity was recorded  simultaneously for the two possible spin polarizations of the incoming electron beam $|+\rangle$ and $|-\rangle$.  This means that first the scattering geometry was adjusted and then the intensity of the scattered electrons was measured after energy analysis. When recording the intensity of the scattered beam two values for the intensity were recorded; one for electrons with $|+\rangle$ spin state and the other one for electrons with $|-\rangle$ spin state. Changing the incoming spin state from $|+\rangle$ to $|-\rangle$ was realized by reversing the helicity of the laser beam used for the excitation and emission of the spin-polarized electrons from the photocathode. The scattered electron beam was energy analyzed without any further spin analysis.

In order to collect the spectra in off-specular geometry at a certain wavevector transfer  $q$ the scattering geometry was adjusted to realize the required wavevector transfer. The in-plane wavevector transfer is given by  $q = k_{i} \sin\theta_i - k_{s} \sin(\theta_{0}-\theta_i)$, where $k_{i}$ ($k_{s}$) is the magnitude of the wavevector of the incident (scattered) electrons, and  $\theta_i$ ($\theta_{0}$)
is the angle between the incident beam and sample normal (the scattered beam). Different wavevectors transfers were achieved by changing the scattering angles, i.e., by rotating the sample about its main axis. In the experiments $\theta_{0}$ was kept at $80^{\circ}$. The spectra were recorded along the $\bar{\Gamma}$--$\bar{\rm{X}}$ direction of the surface Brillouin zone.  The wavevector resolution of the experiment is given by $\Delta q=\sqrt{2mE_{i}}/\hbar[\cos\theta_i
+\cos(\theta_{0}-\theta_i)]\Delta\theta_i$.  $E_i$ denotes the energy of the incident beam and $\Delta\theta_i$ depends of the spectrometer design (in our case $\Delta\theta_i=2^\circ$). The resolution of the spectrometer in momentum space is about 0.03~\AA$^{-1}$.

\subsection{Theory}\label{Sec:Theo}

\subsubsection{Theory of spin dependent scattering cross-section}\label{Sec:scatteringtheo}

We define $|m\rangle$, $|n\rangle$ as many-body states of the sample with energies $E_m$, $E_n$ and $|i\rangle$, $|f\rangle$ as initial and final states of the electron with energies $E_i$, $E_f$. The general definition of the differential scattering cross-section is given by \cite{Berthod2018}
\begin{equation}\label{Eq:crossgeneral}
	\frac{d^2S}{d\Omega d\hbar\omega}=\left(\frac{2\pi}{\hbar}\right)^4m_e^2\frac{k_s}{k_i}\sum_{mn}
	\frac{e^{-E_m/k_{\mathrm{B}}T}}{Z}|\langle n,f|\hat{T}(E_m+E_i)|m,i\rangle|^2\delta(E_m+E_i-E_n-E_f),
\end{equation}
where $m_e$ is the electron mass, $k_i=|\vec{k}_i|$ and $k_s=|\vec{k}_s|$ are the norms of the three-dimensional wavevectors of the incident and scattered electron, respectively, $k_{\mathrm{B}}$ is the Boltzmann constant, $Z=\sum_me^{-E_m/k_{\mathrm{B}}T}$ is the partition function, and $\hbar\omega=E_i-E_f$ is the energy-loss of electrons during the scattering process. $\hat{T}(E)$ is the many-body t-matrix given by $\hat{T}(E)=\hat{V}+\hat{V}(E+i0-\hat{H})^{-1}\hat{T}(E)$ with $\hat{H}$ the Hamiltonian of the sample and $\hat{V}$ the interaction energy of the incident electron with the sample. We use the first-order (Born) approximation, $\hat{T}(E)\approx\hat{V}$.

The interaction energy of the incident electron with the sample is written as $\hat{V}=\hat{V}_{\mathrm{H}}+\hat{V}_{\mathrm{SOC}}$, ignoring the exchange-correlation term. $\hat{V}_{\mathrm{H}}$ is the Hartree energy, i.e., the electrostatic interaction with the charge density in the sample. We define the total charge density operator $\hat{\rho}(\vec{R})$ given in the units of Coulomb per unit volume, where $\vec{R}$ is a three-dimensional position vector. For the given geometry it can be decomposed into the in- and out-of-plane components $\vec{R}=(\vec{r},z)$. Here $z$ represents the coordinate normal to the surface and the sample is placed in the $x$--$y$ plane at $z<0$ (the surface is located at $z=0$). Note that $\hat{\rho}(\vec{R})$ includes both negative as well as positive charges in the sample. This operator acts on the many-body states with matrix elements $\langle n|\hat{\rho}(\vec{R})|m\rangle$. The Hartree energy is
\begin{equation} \label{Hartree}
	\hat{V}_{\mathrm{H}}(\vec{R})=e\int d\vec{R}'\,\frac{\hat{\rho}(\vec{R}')}
	{4\pi\epsilon_0|\vec{R}-\vec{R}'|}\sigma_0,
\end{equation}
where $\sigma_0$ is the identity matrix in spin space, since the Hartree potential conserves the spin of the electron during the scattering process. $\hat{V}_{\mathrm{SOC}}$ is the spin-orbit interaction given by
\begin{equation}
	\hat{V}_{\mathrm{SOC}}(\vec{R})=\frac{e\hbar}{4m_e^2c^2}\vec{\sigma}\cdot
	\left[\hat{\vec{E}}(\vec{R})\times\vec{p}\right].
\end{equation}
Here $\vec{\sigma}$ is the vector of Pauli matrices, $\hat{\vec{E}}(\vec{R})=-\vec{\nabla}\hat{V}_{\mathrm{H}}(\vec{R})/e$ is the electric field due to the charge distribution $\hat{\rho}(\vec{R})$, and $\vec{p}=-i\hbar\vec{\nabla}$ is the momentum operator.

In the next step we introduce the wavefunctions of the incident and scattered electrons, taking into account spin-dependent reflection coefficients
\begin{subequations}\label{eq:psi_new}\begin{align}
	\psi_i(\vec{R})&=N_ie^{i\vec{k}_i\cdot\vec{r}}\left(\vec{i}e^{ik_i^zz}
	+\mathcal{R}\vec{i}e^{-ik_i^zz}\right)\theta(z)\\
	\psi_s(\vec{R})&=N_se^{i\vec{k}_s\cdot\vec{r}}\left(\vec{s}e^{ik_s^zz}
	+\mathcal{R}\vec{s}e^{-ik_s^zz}\right)\theta(z).
\end{align}\end{subequations}
$N_i$ and $N_s$ are normalization factors, $\theta(z)$ denotes the Heaviside step function, $\vec{i}$ and $\vec{s}$ are two-component vectors representing the initial and final states in some spin-1/2 basis (for instance the $\{|+\rangle,|-\rangle\}$ basis), and $\mathcal{R}$ is the reflection matrix expressed in the same basis.
For definiteness, we note that a matrix $\mathcal{R}$ expressed in the usual basis $\{|\uparrow\rangle,|\downarrow\rangle\}$ with the spin quantization axis along $z$ can be rotated to a basis with quantization axis along a direction defined by the polar and azimuthal angles $(\vartheta,\varphi)$ by means of the unitary transformation $\mathcal{U}^{\dagger}\mathcal{R}\mathcal{U}$ with $\mathcal{U}=\begin{pmatrix}\cos\frac{\vartheta}{2}&-\sin\frac{\vartheta}{2}\\ \sin\frac{\vartheta}{2}e^{i\varphi}&\cos\frac{\vartheta}{2}e^{i\varphi}\end{pmatrix}$. The $\{|+\rangle,|-\rangle\}$ basis corresponds to $(\vartheta,\varphi)=(\pi/2,\pi/2)$ (see the Cartesian coordinates in Fig.~\ref{Fig:Spectra}\textbf{a}).
Inserting Eqs.~(\ref{eq:psi_new}) and (\ref{Hartree}) in Eq.~(\ref{Eq:crossgeneral})  we find that the scattering cross-section including only the Hartree term can be written as
\begin{multline}\label{eq:sigmaH}
	\frac{d^2S_{\mathrm{H}}}{d\Omega d\hbar\omega}=\left(\frac{2\pi}{\hbar}\right)^4m_e^2\frac{k_s}{k_i}
 	\left(\frac{e}{2\epsilon_0q}\right)^2(N_sN_i)^2
	\left|\frac{\vec{s}^{\dagger}\!\cdot\vec{i}}{q+iq_z^-}
	+\frac{\vec{s}^{\dagger}\!\cdot(\mathcal{R}\vec{i})}{q+iq_z^+}
	+\frac{(\mathcal{R}\vec{s})^{\dagger}\!\cdot\vec{i}}{q-iq_z^+}
	+\frac{(\mathcal{R}\vec{s})^{\dagger}\!\cdot(\mathcal{R}\vec{i})}{q-iq_z^-}\right|^2 \\
	\times\int_{-\infty}^0dzdz'\,\mathcal{S}(\vec{q},z,z',\omega)e^{-q|z+z'|},
\end{multline}
where $q_z^{\pm}=k_s^z\pm k_i^z$, $\mathcal{S}(\vec{q},z,z',\omega)=\frac{1}{Z}\sum_{mn}e^{-E_m/k_{\mathrm{B}}T}\langle m|\hat{\rho}(-\vec{q},z)|n\rangle\langle n|\hat{\rho}(\vec{q},z')|m\rangle\delta(\hbar\omega+E_m-E_n)$, and $\vec{q}$ is a two-dimensional vector with $|\vec{q}|=q$.

Likewise, the spin-orbit cross-section can be expressed as
\begin{multline}\label{eq:sigmaSOC}
	\frac{d^2S_{\mathrm{SOC}}}{d\Omega d\hbar\omega}=\left(\frac{2\pi}{\hbar}\right)^4m_e^2\frac{k_s}{k_i}
	\left(\frac{e}{2\epsilon_0q}\right)^2(N_sN_i)^2 |\alpha|^2
	\left|(\mathcal{G_+})^{+-}q\left(k_i^x-ik_i^y\right)+(\mathcal{G_-})^{+-}\left(q_x-iq_y\right)ik_i^z\right.\\
	\left.-(\mathcal{G_+})^{-+}q\left(k_i^x+ik_i^y\right)-(\mathcal{G_-})^{-+}\left(q_x+iq_y\right)ik_i^z+
	\left[(\mathcal{G_+})^{++}-(\mathcal{G_+})^{--}\right]\left(q_xk_i^y-q_yk_i^x\right)\right|^2\\
	\times\int_{-\infty}^0dzdz'\,\mathcal{S}(\vec{q},z,z',\omega)e^{-q|z+z'|},
\end{multline}
where ($\mathcal{G_{\pm}}$) is given by
 	\begin{equation}
 		(\mathcal{G_{\pm}})^{\sigma\sigma'}=\frac{s^*_{\sigma}i^{}_{\sigma'}}{q+iq_z^-}
 		\pm\frac{s^*_{\sigma}\left(\mathcal{R}_{\sigma'\sigma'}i_{\sigma'}
 			+\mathcal{R}_{\sigma'\bar{\sigma}'}i_{\bar{\sigma}'}\right)}{q+iq_z^+}
 		+\frac{\left(\mathcal{R}^*_{\sigma\sigma}s^*_{\sigma}
 			+\mathcal{R}^*_{\sigma\bar{\sigma}}s^*_{\bar{\sigma}}\right)i_{\sigma'}}{q-iq_z^+}
 		\pm\frac{\left(\mathcal{R}^*_{\sigma\sigma}s^*_{\sigma}
 			+\mathcal{R}^*_{\sigma\bar{\sigma}}s^*_{\bar{\sigma}}\right)
 			\left(\mathcal{R}_{\sigma'\sigma'}i_{\sigma'}
 			+\mathcal{R}_{\sigma'\bar{\sigma}'}i_{\bar{\sigma}'}\right)}{q-iq_z^-}.
	\end{equation}
Here $\alpha=-i\hbar^2/(4m_e^2c^2)$, $\bar{\sigma}\equiv-\sigma$, and $\mathcal{R}_{\sigma\sigma'}$ represent the reflection coefficients when an electron with a spin $\sigma$ is impinged onto the sample and an electron with spin $\sigma'$ is detected in the final state after the scattering event.

Equations~(\ref{eq:sigmaH}) and (\ref{eq:sigmaSOC}) shall provide a description for the scattering intensities for any possible spin directions of the incoming and scattered beam, when the Hartree and SOC terms are treated separately. However, the quantities $\mathcal{R}_{\sigma\sigma'}$ are not known in practice. In order to overcome this problem simplifications are needed. As it is apparent from Eq.~(\ref{eq:sigmaSOC}) the spin-orbit cross-section is by a factor $\sim|\alpha qk_i|^2$ smaller than the Hartree cross-section and hence its contribution to the intensity may be neglected. The Hartree contribution by itself should, in principle, conserve the spin during the scattering process. In such a scenario no spin asymmetry is expected. In order to account for the spin dependent effects one may assume that the reflection coefficients are spin dependent even for the case of Hartree scattering. In this case one would observe a spin asymmetry. This means that the role of SOC is to break the spin degeneracy and thereby lead to spin-dependent reflection coefficients.

Our analysis showed that, indeed, in this case the asymmetry caused by the Hartree term introduced in Eq.~(\ref{eq:sigmaH}) does not depend, in first approximation, on $q$ and $\hbar\omega$ and is given by (for an extended discussion see Supplementary Note~2)
\begin{equation}\label{eq:asymmetry1}
\frac{I_{|+\rangle}-I_{|-\rangle}}{I_{|+\rangle}+I_{|-\rangle}}
\approx\frac{\mathcal{R}_{++}^2-\mathcal{R}_{--}^2}
{\mathcal{R}_{++}^2+|\mathcal{R}_{+-}|^2+|\mathcal{R}_{-+}|^2+\mathcal{R}_{--}^2}.
\end{equation}
We define the quantities $|\mathcal{R}_{|+\rangle}|^2= |\mathcal{R}_{++}|^2+|\mathcal{R}_{+-}|^2$ and $|\mathcal{R}_{|-\rangle}|^2= |\mathcal{R}_{--}|^2+|\mathcal{R}_{-+}|^2$, which represent the intensity of the scattered electrons when the spin of the incoming beam is of $|+\rangle$ and $|-\rangle$ character, respectively. Such quantities can be extracted from the spin-dependent elastic reflectivity data. The asymmetry strongly depends on the incident (scattered) energy as well as the scattering geometry, since  $|\mathcal{R}_{|+\rangle}|^2$ and $|\mathcal{R}_{|-\rangle}|^2$ depend on these variables.

In our simulations we use exactly the same formalism introduced by Evan and Mills \cite{Evans1972,Mills1975} and implemented by Lucas and \v{S}unji\'{c} \cite{Sunjic1971,Lucas1972,Lambin1990}). In order to account for the spin dependent effects we use reflection coefficients measured by the elastic reflectivity measurements (see below).

\subsubsection{Simulation}\label{Sec:simulation}

Simulation of the spectra were performed by a numerical scheme based on the dipolar scattering theory. To calculate $P^{\mathrm{sl}}(\omega)$ we first calculate the single-loss probability
\begin{equation}\label{Eq:singleloss}
	P(\omega) = \frac{e^2|\mathcal{R}_{| \sigma \rangle}(E_i)|^2}{4\pi\epsilon_0 \hbar v_{\bot}}\frac{4}{\pi^2}
	\iint_{\Omega}d^2q\,\frac{|\vec{q}| (v_{\bot})^3}
	{\left[(\omega-\vec{q} \cdot \vec{v}_{\parallel})^2 + (|\vec{q}| v_{\bot})^2\right]^2}
	\mathfrak{Im}\left[\dfrac{-1}{g(\vec{q},\omega)+1}\right].
\end{equation}
Here $\epsilon_0$ is the vacuum permittivity, $\hbar$ is the reduced Planck constant, $e$ is the electron charge, $v_{\bot}$ ($v_{\parallel}$) represents the perpendicular (parallel) component of the velocity of the incident electron,  $|\mathcal{R}_{| \sigma \rangle}|=|\mathcal{R}_{|+\rangle}|$ or $|\mathcal{R}_{| -\rangle}|$ denotes reflection coefficient of electrons with the incoming spin state $|+\rangle$ or $|-\rangle$ and strongly depends on the incident energy $E_i$. The integration range $\Omega$ denotes the range of  momentum covered by the exit and entrance slits of the monochromator and analyzer. We emphasize that in Eq.~(\ref{Eq:singleloss}) $\vec{q}$ represents the two-dimensional vector of the momentum transfer parallel to the surface. 

The most important entities in Eq.~\eqref{Eq:singleloss} are $|\mathcal{R}_{| \sigma \rangle}(E_i)|$ and the dielectric response function $g(\vec{q},\omega)$. The former is obtained from the experimental elastic reflection measured for different spin directions $|+\rangle$ and $|-\rangle$. For layered systems, such as our case, the latter can be related to the dielectric function of each individual layers $\varepsilon^{p}(q,\omega)$, where $p$ is the layer index (for details see Refs.~\cite{Sunjic1971,Lucas1972,Lambin1990,Lazzari2018}).

Equation~(\ref{Eq:singleloss}) describes only the single-loss probability for an electron having a wavevector $k_i$ and spin $|\sigma\rangle$ to be scattered from a semi-infinite slab system and loose the energy $\hbar\omega$ at $T=0$~K. The multiple scattering events, the elastic peak, and temperature effects were included using the approach introduced by Lucas and \v{S}unji\'{c} \cite{Sunjic1971,Lucas1972,Lambin1990,Lazzari2018}.

We first construct a multi-slab system by considering one unit cell of FeSe on 17 unit cells (about 6.5~nm) of insulating STO on a semi infinite Nb-doped STO(001) (a sketch of the structure is provided in Supplementary Figure~1). The dielectric function of each individual layer is then written in different contributions i.e., (i) a frequency independent background dielectric constant $\varepsilon^{p}_{\infty}$, (ii) a phononic contribution $\varepsilon^{p}_\mathrm{phonon}$, and (iii) an electronic contribution $\varepsilon^{p}_\mathrm{plasmon}$
\begin{equation}
	\varepsilon^{p}(q,\omega) = \varepsilon^{p}_{\infty} + \varepsilon^{p}_\mathrm{phonon}
	+ \varepsilon^{p}_\mathrm{plasmon}.
	\label{eq:dielectric_function}
\end{equation}
We use the literature values of $\varepsilon^{\rm{FeSe}}_{\infty}=15$ and $\varepsilon^{\rm{STO}}_{\infty}=5.7$ for FeSe and STO, respectively \cite{Yuan2012,Gervais1993,Zhou2017,Zhou2016}.

The phononic contribution to the dielectric function of each layer can be expressed in terms of different phonon contributions
\begin{equation}
	\varepsilon^{p}_\mathrm{phonon} = \sum_{j=1}^{m} \frac{Q_j \omega_{\mathrm{TO},j}^2}
	{\omega_{\mathrm{TO},j}^2 -\omega^2 - i\gamma_{\mathrm{TO},j} \omega},
	\label{eq:ph-oscillators-sum}
\end{equation}
where $m$ is the number of all transverse optical (TO) phonon modes with the oscillator strength $Q_j$, which depends on the splitting between TO and longitudinal optical (LO) modes
\begin{equation}
	Q_j = \frac{\varepsilon_\infty}{\omega_{\mathrm{TO},j}^2} \frac{\prod\limits_{l}
	\left(\omega_{\mathrm{LO},l}^2 - \omega_{\mathrm{TO},j}^2 \right)}
	{\prod\limits_{l\neq j}\left(\omega_{\mathrm{TO},l}^2 - \omega_{\mathrm{TO},j}^2 \right)},
	\label{eq:oscillators-strength}
\end{equation}
where $\omega_{\mathrm{TO},j}$ and $\omega_{\mathrm{LO},j}$ denote the frequency of the $j$-th TO and LO phonon modes, respectively. $\gamma_{\mathrm{TO},j}$ and $\gamma_{\mathrm{LO},j}$ represent their corresponding damping.

In addition, we consider a Drude-like term in the dielectric function of each layer, in order to account for the contribution of the charge carriers
\begin{equation}
	\varepsilon^{p}_\mathrm{plasmon} = -\varepsilon_\infty
	\frac{ \omega_{pl}^2-i(\gamma_{pl}-\gamma_0)\omega}{\omega(\omega+i \gamma_0)},
	\label{eq:Drude}
\end{equation}
where $\omega_{pl}$ denotes the plasma frequency associated with the charge carriers and is directly related to the carrier density $n_c$ and carriers' effective mass $m_{\rm eff}$ by $\omega_{pl} =\sqrt{ \frac{n_c e^2}{\varepsilon_\infty \epsilon_0 m_{\rm eff}}}$. The quantities $\gamma_{pl}$ and $\gamma_0$ are the linewidth broadening of the plasmon peak and are determined by the plasmon relaxation time. For $\gamma_{pl} = \gamma_0$ in Eq.~(\ref{eq:Drude}) one arrives at the well-known Drude term.
For the Nb-STO we use $\omega_{pl}=83$~meV, $\gamma_{pl} =75$~meV and $\gamma_0=5$~meV. The values are estimated by extrapolating the values measured by optical techniques at liquid nitrogen temperature to our measurement temperature ($T=15$~K) \cite{Gervais1993,Eagles1996}. The extrapolation is based on the temperature dependence of effective mass as discussed in detail in Ref.~\cite{Collignon2020}. For FeSe ML we use a Drude term with  $\omega_{pl} =334$~meV estimated based on $\omega_{pl} = \sqrt{\frac{n_c e^2}{\varepsilon_\infty \epsilon_0 m_{\rm eff}}}$, assuming $m_{\rm eff}\simeq3 m_e$ and $n_c=0.12$~$e^{-}$/Fe. The damping parameter was  $\gamma_{pl}=\gamma_{0}=270$~meV.

The TO phonon frequencies and their damping as well as $|\mathcal{R}_{| \sigma \rangle}|^2$ serve as the input of the simulations. The values we used for our simulations are provided in Supplementary Table~1. The values of $|\mathcal{R}_{| \sigma \rangle}|^2$ depend on the incident energy. We take the values from Fig.~\ref{Fig:TandEdep}\textbf{b} based on the expression $|\mathcal{R}_{| + \rangle}|^2=\frac{1+A(E_i)}{2}$ and $|\mathcal{R}_{| - \rangle}|^2=\frac{1-A(E_i)}{2}$. For the data shown in Fig.~\ref{Fig:SpectraSimulation}\textbf{b} the values are $|\mathcal{R}_{| + \rangle}|^2(E_i=4~ \rm{eV})=0.555$ and $|\mathcal{R}_{| - \rangle}|^2(E_i=4~ \rm{eV})=0.445$ for those shown in Fig.~\ref{Fig:SpectraSimulation}\textbf{d} they are $|\mathcal{R}_{| + \rangle}|^2(E_i=6~\rm{eV})=0.53$ and $|\mathcal{R}_{| - \rangle}|^2(E_i=6~\rm{eV})=0.47$.

\section*{Data availability}
The datasets generated and/or analyzed during the current study are available from the corresponding author on reasonable request.

\section*{Code availability}
The codes associated with this manuscript are available from the corresponding
author on reasonable request.

\section*{Author contributions}
Kh.Z. initiated the idea of the study, supervised the project, conceived and planned the experiments, analyzed the experimental data, performed the simulations and wrote the paper. D.R. contributed to carrying out the SPHREELS experiments. F.Y., J.J. and W.W. prepared the samples and performed the STM experiments. C.B. carried out the theoretical modeling and derived all the analytical expressions for the spin-dependent scattering cross-sections.

\section*{Competing interests}
The authors declare no competing interests.

\section*{Additional information}
Correspondence and requests for materials should be addressed to Kh.Z (khalil.zakeri@kit.edu).

\section*{Acknowledgements}
Kh.Z. acknowledges funding from the Deutsche Forschungsgemeinschaft (DFG) through the Heisenberg Programme ZA 902/3-1 and ZA 902/6-1 and the DFG Grant No. ZA 902/5-1.  The research of J.J. and W.W. was supported by DFG through Grant No.
Wu 394/12-1. F.Y.  acknowledges funding from the Alexander von Humboldt Foundation. Kh.Z. thanks the Physikalisches Institut for hosting the group and providing the necessary infrastructure. We thank Janek Wettstein and Markus D\"ottling for developing the first version of the simulation code.

\bibliography{Refs}
\twocolumngrid

\end{document}